\documentclass[11pt]{article}
\usepackage{array}
\usepackage{amsmath}
\usepackage{amssymb}
\newcommand{\asup}[1]{\stackrel{\circ}{a}\,\hspace{-2mm}^{#1}}

\begin{document}
\begin{titlepage}
\begin{center}
\hfill hep-th/0705.2347\\
\vskip 1cm
\begin{LARGE}
\textbf{Five-dimensional vector-coupled supergravity on a manifold with boundaries} \footnote{ Work supported
 by the European Commission RTN program ``Constituents, Fundamental Forces and Symmetries of the Universe" MRTN-CT-2004-005104 and by INFN, PRIN prot.2005024045-002}
\end{LARGE}\\
\vspace{1.0cm}
Sean McReynolds \footnote{sean.mcreynolds@mib.infn.it}\\
\vspace{.35cm}
\emph{University of Milano-Bicocca and INFN Milano-Bicocca\\
Piazza della Scienza 3, 20126 Milano, Italy}\\
\vspace{1.0cm} {\bf Abstract}
\end{center}
We consider the bosonic and fermionic symmetries of five-dimensional Maxwell- and Yang-Mills-Einstein supergravity theories on a spacetime with boundaries (isomorphic to $M\times S^{1}/\mathbb{Z}_{2}$).  Due to the appearance of the ``Chern-Simons" term, the classical action is not generally invariant under gauge and supersymmetries.  Once bulk vector fields are allowed to propagate on the boundaries, there is an ``inflow" governed by the rank-3 symmetric tensor that defines the five-dimensional theories.  We discuss the requirements that invariance of the action imposes on new matter content and boundary conditions. 
\vfill {\flushleft {May 2007}}
\end{titlepage}

\section{Introduction}

Theories on $D>4$ dimensional spacetimes with spatially separated boundaries or domain walls are of interest in various scenarios.  They can realize beyond Standard Model scenarios in which some problems of 4D approaches are ameliorated.  They can also be useful for understanding the structure of string/M-theory.  As a prime example, 11D supergravity on a spacetime with boundaries ultimately leads to a description of strongly coupled $E_{8}\times E_{8}$ string theory~\cite{Horava:1996ma}.    

Here we'll consider a spacetime that's topologically $\mathbb{R}^{4}\times S^{1}/\mathbb{Z}_{2}$, where the $\mathbb{Z}_{2}$ acts non-freely on $S^{1}$, corresponding to a spacetime with two spatially separated boundaries.  The boundary conditions on fields arise from lifting the $\mathbb{Z}_{2}$ action from the spacetime to a bundle associated with a symmetry group of the theory.  Once a choice of lifting is made, Kaluza-Klein zero modes of some fields and local symmetries do not exist.  If a 5D theory has gauge and supersymmetries, the 4D theory on the $\mathbb{Z}_{2}$ fixed planes will have half supersymmetry, (generally) a reduced gauge group, and a restricted set of light fields relative to the 5D theory.    

In this paper, we consider $\mathcal{N}=2$ 5D supergravity coupled to $\mathcal{N}=2$ vector multiplets in which some vector fields have massless propagating modes on the boundaries of the spacetime.  In this sense, this is a generalization of previous work such as~\cite{JB1, JB2, offshell1}, and we follow the conventions in~\cite{M05b}. In section~\ref{sec:sugra} we review the framework of 5D supergravity that we use, and the parity assignments of the theory compactified on $S^{1}/\mathbb{Z}_{2}$. In section~\ref{sec:bosonic}, we discuss the anomaly inflow associated with local bosonic transformations, again arising from the `Chern-Simons' term.  This serves as a requirement on chirally coupled fermions in the quantum theory.  In section~\ref{sec:supersymmetry} we consider supersymmetry.  With a simple initial choice of boundary conditions consistent with the parity assignments, the action is not supersymmetric due to the ``Chern-Simons" term of the classical theory.  We discuss the boundary conditions that appear to resolve this.  Basic requirements for the closure of the susy algebra and existence of global Killing spinors are also considered, and are similar to the simple supergravity case in~\cite{JB1}. In the appendix, we include some supplementary details about 5D vector-coupled supergravity.

\section{Supergravity on $M_{4}\times S^{1}/\mathbb{Z}_{2}$}\label{sec:sugra}

We wish to consider five-dimensional $\mathcal{N}=2$ Maxwell-Einstein and Yang-Mills-Einstein supergravity theories (MESGTs and YMESGTs, respectively) on a spacetime that's topologically $\mathbb{R}^{4}\times S^{1}/\mathbb{Z}_{2}$.  An $\mathcal{N}=2$ 5D MESGT is obtained by coupling minimal 5D supergravity to $n_{V}$ abelian vector multiplets~\cite{GST84a}.  The total field content comprises a f\"{u}nfbein, gravitini, vectors, spin-1/2 fields and scalars:
\[\{e^{\hat{m}}_{\hat{\mu}},\,\Psi^{i}_{\hat{\mu}},\, A^{I}_{\hat{\mu}},\,\lambda^{\tilde{p}\,i},\,\phi^{\tilde{x}}\},\]
where $\hat{\mu} = 1,\ldots,5$ is a curved spacetime index; $\hat{m}=\bar{1},\ldots,\bar{5}$ is a flat spacetime index; \;$I=0,\,\ldots ,\,n_{V}$ labels the vectors of the theory (including the ``bare" graviphoton); \;$\tilde{x}=1,\,\ldots,\,n_{V}$ labels the scalars of the vector multiplets (as well as serves as the curved index for the real target space $\mathcal{M}_{R}$ parametrized by those scalars); \;$\tilde{p}=1,\,\ldots,\,n_{V}$ labels the spin-1/2 fields of the vector multiplets (and is a flat index for the scalar manifold); and $i=1,2$ is an index for the $SU(2)_{R}$ (rigid) automorphism group of the superalgebra.  Up to four-fermion terms, the 5D Lagrangian is~\cite{GST84a} 
\begin{equation} \begin{split} 
e^{-1}\mathcal{L}_{(5)}=&-\frac{1}{2\kappa^{2}}R
-\frac{1}{4}\stackrel{\circ}{a}_{IJ}F^{I}_{\hat{\mu}\hat{\nu}}F^{J\;\hat{\mu}\hat{\nu}}
-\frac{3}{4\kappa^{2}}\stackrel{\circ}{a}_{IJ}\partial_{\hat{\mu}}h^{I}\partial^{\hat{\mu}}h^{J} \\
&-\frac{1}{2\kappa^{2}}\bar{\Psi}^{i}_{\hat{\mu}}\Gamma^{\hat{\mu}\hat{\nu}\hat{\rho}}\nabla_{\hat{\nu}}\Psi_{\hat{\rho}\,i}+\frac{\kappa e^{-1}}{6\sqrt{6}}C_{IJK}\epsilon^{\hat{\mu}\hat{\nu}\hat{\rho}\hat{\sigma}\hat{\lambda}}F^{I}_{\hat{\mu}\hat{\nu}}F^{J}_{\hat{\rho}\hat{\sigma}}A^{K}_{\hat{\lambda}}\\
&+\frac{i\sqrt{6}}{4\kappa}\bar{\lambda}^{i\,\tilde{p}}\Gamma^{\hat{\mu}}\Gamma^{\hat{\nu}}\Psi_{\hat{\mu}\,i} h^{\tilde{p}}_{I}\partial_{\hat{\mu}}h^{I} +\frac{1}{4}h^{\tilde{p}}_{I}\bar{\lambda}^{i\,\tilde{p}}\Gamma^{\hat{\mu}}\Gamma^{\hat{\nu}\hat{\rho}}\Psi_{\hat{\mu}\,i}F^{I}_{\hat{\nu}\hat{\rho}}\\
&+\frac{i\kappa}{2\sqrt{6}}\left(\frac{1}{4}\delta_{\tilde{p}\tilde{q}}h_{I}+T_{\tilde{p}\tilde{q}\tilde{r}}h^{\tilde{r}}_{I}\right)\bar{\lambda}^{i\,\tilde{p}}\Gamma^{\hat{\mu}\hat{\nu}}\lambda^{\tilde{q}}_{i}F^{I}_{\hat{\mu}\hat{\nu}}\\
&-\frac{3i}{8\sqrt{6}\kappa}h_{I}\left(\bar{\Psi}^{i}_{\hat{\mu}}\Gamma^{\hat{\mu}\hat{\nu}\hat{\rho}\hat{\sigma}}\Psi_{\hat{\nu}\,i}F^{I}_{\hat{\rho}\hat{\sigma}}+2\bar{\Psi}^{\hat{\mu}\,i}\Psi^{\hat{\nu}}_{i}F^{I}_{\hat{\mu}\hat{\nu}}\right)\\
&-\frac{1}{2}\bar{\lambda}^{i\tilde{p}}\left(\Gamma^{\hat{\mu}}\nabla_{\hat{\mu}}\delta^{\tilde{p}\tilde{q}}-\sqrt{\frac{3}{2\kappa^{2}}}h^{\tilde{x}}_{I}\Omega^{\tilde{p}\tilde{q}}_{\tilde{x}}\Gamma^{\hat{\mu}}\partial_{\hat{\mu}}h^{I}\right)\lambda^{\tilde{q}}_{i},
\end{split} \label{lagrangian}\end{equation}
with supersymmetry transformations
\begin{equation}\begin{split}
\delta_{\epsilon} e^{\hat{m}}_{\hat{\mu}}&=\frac{1}{2}\bar{\epsilon}^{i}\Gamma^{\hat{m}}\Psi_{\hat{\mu}}\\
\delta_{\epsilon} \Psi_{\hat{\mu}i}&=\nabla_{\hat{\mu}}\epsilon_{i}+\frac{i\kappa}{4\sqrt{6}}h_{I}\left(\Gamma^{\hat{\nu}\hat{\rho}}_{\hat{\mu}}-4\delta^{\hat{\nu}}_{\hat{\mu}}\Gamma^{\hat{\rho}}\right)F^{I}_{\hat{\nu}\hat{\rho}}\epsilon_{i}\\
\delta_{\epsilon} A^{I}_{\hat{\mu}}&=-\frac{1}{2}h^{I}_{\tilde{p}}\bar{\epsilon}^{i}\Gamma_{\hat{\mu}}\lambda^{\tilde{p}}_{i}+\frac{i\sqrt{6}}{4}h^{I}\bar{\Psi}^{i}_{\hat{\mu}}\epsilon_{i}\\
\delta_{\epsilon} \lambda^{\tilde{p}}_{i}&=\frac{i}{\kappa}\sqrt{\frac{3}{8}}h^{\tilde{p}}_{I}\partial_{\hat{\mu}}h^{I}\Gamma^{\hat{\mu}}\epsilon_{i}+\frac{1}{4}h^{\tilde{p}}_{I}F^{I}_{\hat{\mu}\hat{\nu}}\Gamma^{\hat{\mu}\hat{\nu}}\epsilon_{i}\\
\delta_{\epsilon} h^{I}&=-\frac{i\kappa}{\sqrt{6}}h^{I}_{\tilde{p}}\bar{\epsilon}^{i}\lambda^{\tilde{p}}_{i},
\end{split}\end{equation}
where $e$ is the f\"{u}nfbein determinant and $\kappa$ is the gravitational coupling.  The MESGT is entirely determined by the form of the rank-3 symmetric tensor $C_{IJK}$.  The $h^{I}$, $h^{I}_{\tilde{x}}$, $h^{I}_{\tilde{p}}$, $T_{\tilde{p}\tilde{q}\tilde{r}}$ and $\Omega^{\tilde{p}\tilde{q}}_{\tilde{x}}$ are functions of the scalars $\phi^{\tilde{x}}$ determined by the condition $\mathcal{V}=C_{IJK}h^{I}h^{J}h^{K}=1$ (see appendix for some useful relations).  
The scalar manifold has a (possibly trivial) group of isometries $Iso(\mathcal{M}_{R})$, which contains a rigid symmetry group $G$ of the Lagrangian (the symmetries of $C_{IJK}$).  The full rigid symmetry group of the Lagrangian is then $G\times SU(2)_{R}$.  The vector fields furnish a representation $R_{V}[G]$ of $G$, which is not necessarily irreducible. 

 A subgroup $K\subset G$ can then be gauged~\cite{Gunaydin:1984ak,Gunaydin:1984nt,Gunaydin:1999zx,Ceresole:2000jd}, yielding a YMESGT, if the adjoint representation appears in the decomposition
\[\mathbf{R_{V}}[G]=\mbox{\textbf{adj}}[K]\oplus \mbox{\textbf{non-singlets}}[K] \oplus \mbox{\textbf{singlets}}[K].\]   
Any non-singlet vector fields must be dualized to tensor fields~\cite{GZ99} satisfying a first order field equation such that the degrees of freedom remain the same~\cite{selfdual}.  $K$-singlets are also called spectators since nothing is charged with respect to them.  When $K$ is compact, the graviphoton is a spectator.  For simplicity, we will ignore tensor couplings and spectator vector fields (other than the graviphoton if it is one) in this paper.  Derivatives are then made $K$-covariant, and the abelian fieldstrengths of the adjoint vectors are replaced by non-abelian ones:
\[ \begin{split}
F^{I}_{\hat{\mu}\hat{\nu}}&\longrightarrow \mathcal{F}^{I}_{\hat{\mu}\hat{\nu}}=F^{I}_{\hat{\mu}\hat{\nu}}+gf^{I}_{JK}A^{J}_{\hat{\mu}}A^{K}_{\hat{\nu}}\\
\partial_{\hat{\mu}}h^{I}&\longrightarrow D_{\hat{\mu}}h^{I}=\partial_{\hat{\mu}}h^{I}+gA^{J}_{\hat{\mu}}f^{I}_{JK}h^{K}\\
\nabla_{\hat{\mu}}\lambda^{i\,\tilde{p}}&\longrightarrow D_{\hat{\mu}}\lambda^{i\,\tilde{p}}=\nabla_{\hat{\mu}}\lambda^{i\,\tilde{p}}+gA^{I}_{\hat{\mu}}L^{\tilde{p}\tilde{q}}_{I}\lambda^{i\,\tilde{q}}.
\end{split} \]
The exception is the abelian ``Chern-Simons" $FFA$ term of the MESGT, which is replaced by
\begin{equation} \begin{split}
\frac{\kappa e^{-1}}{6\sqrt{6}}C_{IJK}\,\epsilon^{\hat{\mu}\hat{\nu}\hat{\rho}\hat{\sigma}\hat{\lambda}}\{ &F^{I}_{\hat{\mu}\hat{\nu}}F^{J}_{\hat{\rho}\hat{\sigma}}A^{K}_{\hat{\lambda}}+\frac{3}{2}gF^{I}_{\hat{\mu}\hat{\nu}}A^{J}_{\hat{\rho}}\,(f^{K}_{LM}A^{L}_{\hat{\sigma}}A^{M}_{\hat{\lambda}}) \\
+ &\frac{3}{5}g^{2}(f^{J}_{GH}A^{G}_{\hat{\nu}}A^{H}_{\hat{\rho}})(f^{K}_{LF} A^{L}_{\hat{\sigma}}A^{F}_{\hat{\lambda}})\,A^{I}_{\hat{\mu}}\},\label{CS}
\end{split} \end{equation}
where the $F^{I}_{\mu\nu}$ are still abelian fieldstrengths.
The tensor $C_{IJK}$ is a rank-3 symmetric invariant of the subgroup $K\subset G$.  In addition, a new Yukawa term, whose form is irrelevant here, must also be added to the Lagrangian.  

We can now consider a MESGT or YMESGT on $M_{4}\times S^{1}/\mathbb{Z}_{2}$.  This can be viewed either as a spacetime with boundaries (the ``downstairs picture"), or by compactifying on the covering space $M_{4}\times S^{1}$, assigning $\mathbb{Z}_{2}$ parities consistently to the fields and other objects in the theory, and identifying spacetime points under $x^{5}\rightarrow -x^{5}$ (the ``upstairs picture").  We coordinatize the $S^{1}$ covering space as $[-\pi R,\,\pi R]$, where $\{-\pi R\}\equiv \{+\pi R\}$.  The fixed points of the $\mathbb{Z}_{2}$ action on the circle are then $\{0\},\{\pi R\}$.

Across the $\mathbb{Z}_{2}$ fixed-points, fields can generally satisfy a jump condition~\cite{Bagger:2001qi} $\Phi(x^{5}_{fp}-\xi^{5})=U \Phi(x^{5}_{fp}+\xi^{5})$, where $x^{5}\pm \xi^{5}$ is in an open neighborhood of $x^{5}$, and $U$ is a representation of the $\mathbb{Z}_{2}$ action on the space of fields $\Phi(x,x^{5})$.  We consider fields to be of the form
\begin{equation}
\Phi=\Phi_{c}+\aleph\,\theta(x^{5})\Phi_{\theta}\label{fieldexp}
\end{equation}
where $\theta(x^{5})$ is $-1$ for $(-\pi R, 0)$ and $+1$ for $(0,+\pi R)$.  $\Phi_{c}$ carries the same parity as $\Phi$, while $\Phi_{\theta}$ has opposite parity of $\Phi$.  Both $\Phi_{c}$ and $\Phi_{\theta}$ are continuous but not necessarily smooth functions (they are generally $C^{0}$).  Odd fields will not have independent propagating modes on the fixed planes, nor independent zero modes in the full spacetime.  For example, consider an odd field satisfying a free equation of motion on $S^{1}$; then $\Phi_{c}=\Sigma_{n}\Phi_{n}(x^{\mu})\sin(nx^{5}/R)$ and $\Phi_{\theta}=\Sigma_{n}\Phi_{n}(x^{\mu})\cos(nx^{5}/R)$.  

Let $\hat{\mu}=(\mu,5)$, where $\mu$ is a 4D curved spacetime index.  For 5D scalar fields and anti-symmetric tensor fields with only 4D indices, $G_{\mu_{1}\cdots\mu_{n}}$, satisfying 2nd order field equations, $\theta(x^{5})$ would introduce squared Dirac distributions $\delta^{2}(x^{5})$ in the Lagrangian ($\delta(0)$ in the action).  Therefore, the truncation $\aleph=0$ is typically made for these.  On the other hand, we can allow odd fermionic fields, anti-symmetric tensor fields $G_{5\nu_{1}\cdots\nu_{m}}$ satisfying 2nd order field equations, and anti-symmetric tensor fields $G_{\mu_{1}\cdots\mu_{n}}$ satisfying 1st order field equations to jump at fixed points.  We are regarding the f\"{u}nfbein or 5D spin connection as fundamental fields here, so the above applies to them as rank-1 tensor fields.  In the downstairs picture all fields are expected to satisfy well-defined boundary conditions. 

The field content on $M_{4}\times S^{1}$ becomes 
\[\{(e^{m}_{\mu},C_{\mu},e^{\sigma}),(\Psi^{i}_{\mu},\psi^{i}),(A^{I}_{\mu},A^{I}),\lambda^{\tilde{p}\,i},h^{I}\},\]  
where $e^{\bar{5}}_{5}=e^{\sigma}$ and $e^{\bar{5}}_{\mu}\propto C_{\mu}$. The consistent parity assignments for the generally coupled theory can be found in~\cite{M05b}.  Working in a canonical basis in which $I=0$ labels the `bare' 5D graviphoton, we make the index splitting $I=(0,\alpha,a)$ corresponding to the following parity assignments for the bosonic fields\footnote{Odd $h_{\alpha}$ and $\stackrel{\circ}{a}_{\alpha a}$ do not vanish at fixed points if $C_{\alpha a b}\neq 0$ for values of $\alpha$ that are `spectators' of the gauge group $K$ (or rigid group $G$).} 
\begin{center}
\begin{tabular}{|c|c|}  \hline
Even & Odd \\ \hline 
$g_{\mu\nu}\;\;e^{\sigma}$ & $C_{\mu}$\\ 
$A^{\alpha}_{\mu}\;\;A^{0}\;\;A^{a}$ & $A^{a}_{\mu}\;\;A^{0}_{\mu}\;\;A^{\alpha}$ \\
$h^{a}\;\;h^{0}$ & $h^{\alpha}$ 
 \\\hline
\end{tabular}
\end{center} 
The value $\wp$ in $\alpha=1,\ldots,\wp$ and $a=\wp+1,\ldots, n_{V}$ is arbitrary here. 

In terms of 2-component spinors (see appendix) the fermions and susy parameters are
\begin{equation}
\lambda^{\tilde{p}\,1}=\left(\begin{array}{c}
\delta^{\tilde{p}}\\
e\gamma^{\tilde{p}\,*}
\end{array}\right)\;\;\;\;\lambda^{\tilde{p}\,2}=\left(\begin{array}{c}
\gamma^{\tilde{p}}\\
-e \delta^{\tilde{p}\,*}
\end{array}\right)
\label{spinors1}\end{equation}
\[
\Psi^{1}_{\hat{\mu}}=\left(\begin{array}{c}
\alpha_{\hat{\mu}}\\
e\beta^{*}_{\hat{\mu}}
\end{array}\right)\;\;\;\;\Psi^{2}_{\hat{\mu}}=\left(\begin{array}{c}
\beta_{\hat{\mu}}\\
-e \alpha^{*}_{\hat{\mu}}
\end{array}\right);\;\;\;\;
\epsilon^{1}=\left(\begin{array}{c}
\eta\\
e\zeta^{*}
\end{array}\right)\;\;\;\;\epsilon^{2}=\left(\begin{array}{c}
\zeta\\
-e \eta^{*}
\end{array}\right),
\]
Splitting $\tilde{p}=(p,\rho)$, $\rho=1,\ldots,\wp$ and $p=\wp+1,\ldots,n_{V}$, the parity assignments are
\begin{center}
\begin{tabular}{|c|c|}  \hline
Even & Odd \\ \hline 
$\delta^{\rho}\;\;\gamma^{p}$ & $\delta^{p}\;\;\gamma^{\rho}$\\ 
$\alpha_{\mu}\;\;\beta_{5}$ & $\alpha_{5}\;\;\beta_{\mu}$ \\
$\eta$ & $\zeta$ 
 \\\hline
\end{tabular}
\end{center} 
The parities of the $C_{IJK}$ as well as the parameters $\alpha^{I}$ and structure constants $f^{I}_{JK}$ of the rigid or gauge algebra, are
\begin{center}\begin{tabular}{|c|c|}  \hline
Even & Odd \\ \hline  
$C_{abc}\;\;C_{a\alpha\beta}$ & $C_{\alpha ab}\;\;C_{\alpha\beta\gamma}$ \\
$f^{\alpha}_{\beta\gamma}\;\;f^{\alpha}_{ab}$ & $f^{a}_{bc}\;\;f^{a}_{\alpha\beta}$\\
$f^{0}_{a \beta}\;\;f^{0}_{\alpha 0}$ & $f^{0}_{\alpha\beta}$\\
$\alpha^{\beta}$ & $\alpha^{0}\;\;\alpha^{b}$\\ \hline
\end{tabular}\end{center}
where $f^{I}_{JK}$ vanishes if any of the indices correspond to 5D spectator vector fields; and permutations of the indices have the same parity.  Odd $C_{IJK}$ and $f^{I}_{JK}$ can be redefined as $\theta(x^{5})f^{I}_{JK}$ and $\theta(x^{5})C_{IJK}$. 

\section{Bosonic symmetries}\label{sec:bosonic}    

\subsection{Symmetry algebras}\label{sec:susyalgebra}

In a 5D MESGT, the rigid symmetry algebra $\mathfrak{g}$ of the Lagrangian acts on vector multiplet scalars via 
\[
\delta_{\Lambda} h^{I}= f^{I}_{JK}\Lambda^{J}h^{K}.
\]
However, on $M_{4}\times S^{1}/\mathbb{Z}_{2}$ only the subalgebra $\mathfrak{g}_{\alpha}$ parametrized by the $\Lambda^{\alpha}$ are rigid 5D symmetries.  It appears in the reductive homogeneous decomposition $\mathfrak{g}=\mathfrak{g}_{\alpha}\oplus \mathfrak{t}^{a}$, for which the $h^{\alpha}$ and $h^{a}$ furnish a representation.  However, as in the case of dimensional reduction~\cite{GST84a, Cremmer:1984hj}, the scalar isometries are enlarged on the fixed planes so that the manifest symmetry subalgebra of the Lagrangian is
\[
(\mathfrak{g}_{\alpha}\oplus\beta\oplus \mathfrak{t}^{\tilde{I}})\,\circledS\,\mathfrak{t}^{\hat{a}},\label{fpalg}
\]
where $\circledS$ is a semidirect sum. The indices $0,a$ have been grouped into the $\mathfrak{g}_{\alpha}$-singlet index $\tilde{I}=(0,\tilde{a})$ and non-singlet index $\hat{a}$. The $\mathfrak{t}^{a}$ are constant shift symmetries of the scalars $A^{a}$, and $\beta$ are dilations.\footnote{In the dimensional reduction, the translation algebra also acts on the fieldstrengths $F^{I}_{\mu\nu}$ of the 4D theory, but this doesn't carry over to the fixed plane theories since it involves the Kaluza-Klein vector $C_{\mu}$.} 

By contrast, the symmetry transformations of a YMESGT with gauge algebra $k$ are all consistent as 5D symmetries on $M_{4}\times S^{1}/\mathbb{Z}_{2}$. The symmetry algebra at the fixed points is the reductive homogeneous space 
\begin{equation}
k_{\alpha}\oplus \mathfrak{p}^{a},
\label{fpalgebra}\end{equation}
where $k_{\alpha}$ is the surviving gauge algebra, and $\mathfrak{p}^{a}$ act as local shifts of the KK-exicted modes of $A^{a}$, so do not survive as symmetries of the 4D effective Lagrangian.  The would-be constant shift symmetries $\mathfrak{t}^{\hat{a}}$ are broken by gauge couplings. 

\subsection{Anomalies}\label{sec:anomalies}

Five-dimensional theories on a spacetime with boundaries can have pure gauge or mixed anomalies due to the presence of fermions in complex representations of the boundary gauge group, $K_{\alpha}$.  Discussion of such anomalies in orbifold theories can be found in~\cite{AHCG01, anomalies}.    
In the present case (i) charged chiral multiplets coming from 5D vector or tensor multiplets appear in real representations of $K_{\alpha}$; (ii) the chiral multiplet coming from the 5D supergravity multiplet is a $K_{\alpha}$-singlet (since the gauge group is contained in the isotropy group of the real scalar manifold $K_{\alpha}\subset Iso_{0}(M_{R})$, and the 5D sugra multiplet is a singlet of $Iso_{0}(M_{R})$);    
and (iii) the 4D spin-3/2 fermion is in the 4D supergravity multiplet and will not have anomalous gauge couplings if we do not gauge a subgroup of the $SU(2)_{R}$ automorphism group.  (Furthermore, since $R$-symmetries are not gauged, there aren't any Fayet-Illiopoulos terms.)  Therefore, the only bulk fermions that may have anomalous gauge couplings in the quantum theory sit in the chiral multiplets coming from charged 5D hypermultiplets.  There may also be new chirally coupled massless fermions supported on the boundaries, though we do not assume this \textit{a priori}.  

In five dimensional domain wall or $S^{1}/\mathbb{Z}_{2}$ backgrounds, it has been shown that if bulk fermions appear in anomaly-free representations from a 4D point of view, the 4D effective quantum theory will be anomaly free~\cite{AHCG01}.  Any anomaly incurred is located at the fixed points (or domain walls), the form of the bulk wavefunctions being irrelevant.  In the case where bulk fermions have a mass term, e.g. from scalar vevs, the freedom from anomalies can be seen as a cancellation between a quantum anomaly and a 1-loop Chern--Simons (CS) inflow at each wall or fixed plane~\cite{AHCG01}.  The CS term has $\mathbb{Z}_{2}$ odd parity across the wall or fixed plane, associated with the $\mathbb{Z}_{2}$-odd mass parameter, which in turn is often taken to be a vev of a scalar field. 
 
Due to chiral couplings of fermions arising from bulk $\mathcal{N}=2$ hypermultiplets or boundary $\mathcal{N}=1$ chiral multiplets, the gauge variation of the effective quantum action therefore includes
\begin{equation}
\delta_{\alpha} S_{eff}\propto D_{\alpha\beta\gamma}\epsilon^{\mu\nu\rho\sigma}\int_{\partial M} \mathcal{F}^{\alpha}_{\mu\nu}\mathcal{F}^{\beta}_{\rho\sigma}\alpha^{\gamma},
\label{quantanom}\end{equation}
where $D_{\alpha\beta\gamma}=tr_{(f)}[t_{\alpha}\{t_{\beta},t_{\gamma}\}]$ involves the trace over massless left-chiral fermionic species in their representations; and $\mathcal{F}_{\mu\nu}$ denotes the gauge fieldstrengths, whether abelian or non-abelian.  The coefficient of proportionality depends on where the matter is coming from in the model.        

In addition to these quantum anomalies, 5D supergravity has a classical ``Chern-Simons" term, which can contribute to reducible and irreducible inlow anomalies at 4D boundaries due to a non-zero gauge variation~\cite{CH85} (for analogous phenomena in the eleven-dimensional case, see~\cite{Witten:2001uq}).  The Chern-Simons contribution to the action of five-dimensional Maxwell-Einstein supergravity theory is
\[
 S_{CS}=\int_{M}\frac{\kappa}{6\sqrt{6}}\epsilon^{\hat{\mu}\hat{\nu}\hat{\rho}\hat{\sigma}\hat{\lambda}}C_{IJK}F^{I}_{\hat{\mu}\hat{\nu}} F^{J}_{\hat{\rho}\hat{\sigma}} A^{K}_{\hat{\lambda}}, 
\]
where $C_{IJK}$ is a rank-3 symmetric invariant of the rigid group $G$.  On a spacetime that's topologically $\mathbb{R}^{5}$, the full action is invariant under local abelian transformations $\delta_{\alpha}A^{I}_{\hat{\mu}}=\partial_{\hat{\mu}}\alpha^{(I)}$.\footnote{These are not proper ``gauged supergravities", which arise when R-symmetries or scalar isometries are gauged.  We use $\alpha^{(I)}$ for the set of local abelian parameters in a MESGT, and $\alpha^{I}$ for the parameters of $K$ in a YMESGT.}  If we now consider a manifold with boundary, the variation of the action under these abelian transformations yields
 \begin{equation} 
\delta_{\alpha}S_{CS}=\frac{\kappa}{6\sqrt{6}}\epsilon^{\mu\nu\rho\sigma}C_{\alpha\beta\gamma}\int_{\partial{M}}F^{\alpha}_{\mu\nu}F^{\beta}_{\rho\sigma}\alpha^{(\gamma)}, 
\label{anomaly}\end{equation}    
where $C_{\alpha\beta\gamma}$ is a rank-3 symmetric invariant of the rigid symmetry subgroup $G_{\alpha}\subset G$ of the Lagrangian.\footnote{In the upstairs picture, the CS term is $\mathbb{Z}_{2}$ invariant; in the variation of the action, integration by parts results in a fixed-point localized anomaly.} Since the boundaries are oppositely oriented, the flux of abelian currents coming from one boundary is received by the other (i.e., the anomaly globally cancels).  However, there is a classical inflow anomaly at the individual boundaries so that the corresponding local abelian symmetries of a 5D MESGT are broken.  This inflow must vanish or be compensated locally for consistency of the theory.

However, the classical Lagrangian is otherwise invariant under the local abelian transformations we are discussing.  The fermions propagating on the boundaries will not have a chiral anomaly contribution since they are not charged with respect to any of the abelian fields.  Therefore, in dealing with MESGTs in the presence of boundaries, only the theories with $C_{\alpha\beta\gamma}=0$ are invariant under the full set of local abelian transformations.  For theories with $C_{\alpha\beta\gamma}\neq{0}$, we need a quantum anomaly due to boundary fermions charged with respect to the $\wp$ abelian fields $F^{\alpha}_{\mu\nu}$; in that case the $D_{\alpha\beta\gamma}$ in~(\ref{quantanom}) are products of (total) charges of the left-chiral fermions on a given boundary.      

As a set of examples, consider the MESGTs based on Lorentzian Jordan algebras~\cite{Gunaydin:2003yx} $J^{\mathbb{C}}_{(1,N)}$ (represented as matrices over the complex numbers that are Hermitian with respect to a Minkowski metric).  The rigid symmetry group of the theory on $\mathbb{R}^{5}$ is $G=SU(N,1)$ with all the vector fields, including the graviphoton, forming the adjoint representation.  The $C_{IJK}$ are proportional to the $d$-symbols of $SU(N,1)$, so the action contains the abelian Chern-Simons 5-form
\[ \begin{split}
\mathcal{S}_{CS}=\frac{D\,\kappa}{6\sqrt{6}}\int\epsilon^{\hat{\mu}\hat{\nu}\hat{\rho}\hat{\sigma}\hat{\lambda}}\mbox{Tr}[F_{\hat{\mu}\hat{\nu}}F_{\hat{\rho}\hat{\sigma}} A_{\hat{\lambda}}], 
\end{split}\]
where $A_{\hat{\mu}}=t_{I}A^{I}_{\hat{\mu}}$, with $t_{I}\in \mathfrak{su}(N,1)$, and $D\,\mbox{Tr}[t_{I}\{t_{J},t_{K}\}]=C_{IJK}$, with $D\neq 0$ and the trace taken in the adjoint representation.  We now choose parity assignments such that $\alpha$ is $\mbox{adj}[SU(N)]$-valued.  Then on $M\times S^{1}/\mathbb{Z}_{2}$, the rigid symmetry group of the 5D action is $G_{\alpha}=SU(N)$.  Under the \textit{local abelian} transformations $\delta A^{I}_{\hat{\mu}}=\partial_{\hat{\mu}}\alpha^{(I)}$, the variation of the above term yields
\begin{equation} 
\delta S_{CS}=\frac{D\kappa}{6\sqrt{6}}d_{\alpha\beta\gamma} \epsilon^{\mu\nu\rho\sigma}\int_{\partial M} F^{\alpha}_{\mu\nu}F^{\beta}_{\rho\sigma}\,\alpha^{(\gamma)}.\label{inflow}
\end{equation}  
Since $D\neq 0$, we must compensate this as discussed in the previous paragraph.  
 	        
For a YMESGT in which the bulk gauge group $K\subset G$ is broken on the boundaries to $K_{\alpha}$, the susy variation is the same as~(\ref{anomaly}) with $F^{\alpha}_{\mu\nu}\rightarrow \mathcal{F}^{\alpha}_{\mu\nu}$ and $\alpha^{(\alpha)}\rightarrow \alpha^{\alpha}$ being the fieldstrengths and parameters of $K_{\alpha}$.  Therefore, there will be an anomaly inflow of local $K_{\alpha}$ currents when $C_{\alpha\beta\gamma}\neq 0$. The inflow must be compensated by quantum anomalies due to an appropriate set of either bulk hypermultiplets or boundary-supported fields chirally coupled to the $K_{\alpha}$ gauge fields.    
 
As a set of examples, begin with the MESGTs of the previous example.  Then consider the YMESGTs obtained by gauging the rigid symmetry group $K\simeq G=SU(N,1)$ of the MESGT (on $\mathbb{R}^{5}$)~\cite{Gunaydin:2003yx}.  The 5D action on $M_{4}\times S^{1}/\mathbb{Z}_{2}$ has this as a symmetry up to inflow.  The non-abelian Chern-Simons 5-form now appears
\[ \begin{split}
S_{CS}=
\int\frac{D\,\kappa}{6\sqrt{6}\,g^{3}}\epsilon^{\hat{\mu}\hat{\nu}\hat{\rho}\hat{\sigma}\hat{\lambda}}\mbox{Tr}[F_{\hat{\mu}\hat{\nu}}F_{\hat{\rho}\hat{\sigma}} A_{\hat{\lambda}}
+\frac{3}{2}F_{\hat{\mu}\hat{\nu}} A_{\hat{\rho}} [A_{\hat{\sigma}},A_{\hat{\lambda}}]+\frac{3}{5}A_{\hat{\mu}} [A_{\hat{\nu}},A_{\hat{\rho}}] [A_{\hat{\sigma}},A_{\hat{\lambda}}]],
\end{split}\]
where again the trace is in $\mbox{adj}[SU(N,1)]$, and we've rescaled $A^{I}_{\hat{\mu}}\rightarrow gA^{I}_{\hat{\mu}}$, with $g$ the 5D gauge coupling.  The index $\alpha$ can at most be $\mbox{adj}[SU(N)]$-valued~\cite{M05b}, which we take here.
Under gauge transformations connected to the identity, the action has the variation~(\ref{inflow}) with $F^{\alpha}_{\mu\nu}\rightarrow \mathcal{F}^{\alpha}_{\mu\nu}$ and $\alpha^{(\gamma)}\rightarrow \alpha^{\gamma}$.  We must include matter resulting in 4D fermions in appropriate $\mathbb{C}$-representations of $SU(N)$.    
Strictly speaking, this example again requires the addition of boundary-supported fermions such that~(\ref{quantanom}) compensates the non-abelian version of~(\ref{inflow}).  However, there may be 5D theories coupled to hypermultiplets that satisfy this.  In the case of $K_{\alpha}=SU(N)$ in the previous example, we can couple $n$ hypermultiplets with homogeneous scalar manifold $SU(nN,2)/(SU(nN)\times SU(2)\times U(1))$.\footnote{The gauging of $SU(N,1)$ isometries of the quaternionic manifold necessarily involves gauging R-symmetries, which we have not considered in this paper.}  Since the real scalars form the $(nN,2)\oplus(\overline{nN},2)$ of the $SU(nN)\times SU(2)$ subgroup, the orbifold theory will give $n$ left-chiral multiplets in the $\mathbf{N}$ of $SU(N)$ (along with their right-chiral conjugates in the $\mathbf{\bar{N}}$)~\cite{M05a}.  However, from~\cite{M05a} one can see that when other homogeneous hypermultiplet scalar manifolds do admit left-chiral fermions in complex representations, they are generally in complete representations of groups that don't admit complex representations.  Therefore, they cannot contribute to the quantum anomaly.  By compactifying on $S^{1}/(\mathbb{Z}_{2}\times\mathbb{Z}_{2})$, one can make these \textit{incomplete} representations of those groups, and so allow for more options in compensating the classical anomaly inflow.  

\section{Supersymmetry}\label{sec:supersymmetry}

In the orbifold field theory, only half of the massless fermionic spectrum remains on the fixed planes, so we expect half the susy to be broken.  However, the invariance of the action under local susy transformations and the preservation of susy currents must be checked.  The variation of a susy Lagrangian generally involves total spacetime derivatives so it's possible in the presence of a boundary that the action has a non-zero variation, which would require some modification of the theory.  In the upstairs picture, such an ``anomalous susy inflow" arises from integration by parts involving terms in the action with $\theta(x^{5})C^{I_{1}\cdots I_{n}}_{J_{1}\cdots J_{m}}$ factors (where $C$ is some object in the MESGT or YMESGT carrying $G$ or $K$-indices, resp.).         
\subsection{Downstairs picture}\label{sec:downstairs}
Let's first consider MESGTs or YMESGTs with $K_{\alpha}$ abelian, so that only abelian fieldstrengths $F^{\alpha}_{\hat{\mu}\hat{\nu}}$ appear in the following. Imposing simple boundary conditions consistent with $\mathbb{Z}_{2}$ parity assignments
\begin{equation}\left. \begin{split}
&e^{\bar{5}}_{\mu}=0=e^{m}_{5}\\
&h^{\alpha}=0=h^{\alpha}_{p},\;\;\;\;F^{\alpha}_{\mu 5}=0=F^{a}_{\mu\nu}\\
&\delta^{p}=0,\;\;\;\;\;\;\;\;\;\;\;\;\;\;  \gamma^{\rho}=0 \\
&\beta_{\mu}=N\alpha_{\mu},\;\;\;\;\;\;\;\; \alpha_{5}=N\beta_{5}\\
&\zeta=N \eta \end{split}\right\} \;\; \mbox{on}\;\partial M 
\label{bcs}\end{equation}
where $N$ is a non-negative real constant, the downstairs action is invariant up to the topological term (ignoring four-fermion terms)
\begin{equation}
e^{-1}\delta_{\epsilon}S= -\frac{2}{3\sqrt{6}}\int_{\partial M}\kappa e^{-1}\,C_{\alpha\beta\gamma}\epsilon^{\mu\nu\rho\sigma}F^{\alpha}_{\mu\nu}A^{\beta}_{\rho}(\delta_{\epsilon}A^{\gamma}_{\sigma}),\label{susyvar}
\end{equation} 
where 
\[\delta_{\epsilon}A^{\alpha}_{\mu}=-\frac{1}{2}h^{\alpha}_{\tilde{p}}\bar{\epsilon}^{i}\Gamma_{\mu}\lambda^{\tilde{p}}_{i}\;\;\;\; \mbox{on}\; \partial M.\]

However, as mentioned in section 2, the upstairs picture allows some odd bosonic fields to jump at fixed points, corresponding to non-zero boundary conditions in the downstairs picture.  We can therefore look for different boundary conditions such that the `susy inflow' is compensated.
The susy variation of the action contains the terms 
\begin{equation}\begin{split}
e^{-1}\delta_{\epsilon} S\sim \int_{\partial M} & \left[-\stackrel{\circ}{a}_{\alpha\beta}F^{\alpha\,5\mu}(\delta_{\epsilon} A^{\beta}_{\mu})-\frac{1}{2}\bar{\lambda}^{i\,\tilde{p}}\Gamma^{5}(\delta_{\epsilon} \lambda^{\tilde{p}}_{i})\right]\\
&+\int_{M}\frac{1}{4}h^{\tilde{p}}_{\alpha}\bar{\lambda}^{i\,\tilde{p}}\Gamma^{5}\Gamma^{\hat{\mu}\hat{\nu}}(\delta_{\epsilon} \Psi_{5\,i})F^{\alpha}_{\hat{\mu}\hat{\nu}}
+ e^{-1}\delta_{\epsilon} S_{CS},
\end{split}\label{totalvar}\end{equation}
where $\delta_{\epsilon} S_{CS}$ is given by~(\ref{susyvar}).
Using integration by parts to get a bare $\epsilon_{i}$ in the third term, the variation can be written as the two boundary terms
\[
e^{-1}\delta_{\epsilon} S\sim -\frac{3}{4}\int_{\partial M}h^{\alpha\,\rho}\,\bar{\epsilon}^{i}\Gamma_{\mu}\lambda^{\rho}_{i}F^{\mu 5}_{\alpha}+e^{-1}\delta_{\epsilon} S_{CS}.
\]
This vanishes when we impose the boundary condition 
\begin{equation}
F^{\mu 5}_{\alpha}= -\frac{4}{9\sqrt{6}}\kappa e^{-1}\epsilon^{\mu\nu\rho\sigma}C_{\alpha\beta\gamma}F^{\beta}_{\nu\rho}A^{\gamma}_{\sigma}\;\;\;\;\mbox{on}\;\;\partial M.\label{bc}
\end{equation}
The covariant form is related by $F^{\alpha}_{\mu 5}=F^{\nu 5}_{\beta}g_{\mu\nu}g_{55}\asup{\alpha\beta}$, where $g_{\hat{\mu}\hat{\nu}}$ is the 5D metric.  Note that the $\epsilon^{\mu\nu\rho\sigma}C_{\alpha\beta\gamma}F^{\beta}_{\nu\rho}A^{\gamma}_{\sigma}$ are analogs of the abelian Chern-Simons class, valued in the Lie algebras $\mathfrak{g}$ or $\mathbf{k}$ of the groups $G$ or $K$, resp.

When we consider a YMESGT with non-abelian gauge group $K_{\alpha}$, the analysis is the same with $F^{\alpha}_{\hat{\mu}\hat{\nu}}\rightarrow \mathcal{F}^{\alpha}_{\hat{\mu}\hat{\nu}}$ in~(\ref{totalvar}), and with~(\ref{susyvar}) replaced by the variation of~(\ref{CS}). The required boundary condition becomes 
\begin{equation}
\mathcal{F}^{\mu 5}_{\alpha}= -\frac{4}{9\sqrt{6}}\frac{\kappa e^{-1}}{g}\epsilon^{\mu\nu\rho\sigma}C_{\alpha\beta\gamma}\left[F^{\beta}_{\nu\rho}A^{\gamma}_{\sigma}+ \frac{3}{4}A^{\beta}_{\nu}(A^{\alpha'}_{\rho}f^{\gamma}_{\alpha'\beta'}A^{\beta'}_{\sigma})\right]\;\;\;\;\mbox{on}\;\;\partial M,
\label{nonabelianbc}\end{equation}
where we've rescaled $gA^{I}_{\hat{\mu}}\rightarrow A^{I}_{\hat{\mu}}$.  
Note that unlike the abelian case, this does not involve Chern-Simons analogs (which would have a factor of $2/3$ in the second term above).   

\subsection{Upstairs picture}\label{sec:upstairs}
For simplicity, let's again first consider the susy variation of a MESGT or YMESGT with abelian $K_{\alpha}$.  The `Chern-Simons' term now has a contribution with coefficient $\theta(x^{5})C_{\alpha\beta\gamma}$.  When obtaining bare $\epsilon^{i}$ in the variation, integration by parts picks up $\partial_{5}\theta(x^{5})$ yielding a fixed-plane contribution in the action as in~(\ref{susyvar}).  Compensation now occurs if we have the modified field\footnote{In the upstairs picture, we integrate over $S^{1}$ with a factor of $1/2$ to correct for $\mathbb{Z}_{2}$ identification. Also, $\partial_{5}\theta(x^{5})=2\left\{\delta(x^{5})-\delta(x^{5}-\pi R)\right\}\equiv 2\{\delta_{0}-\delta_{\pi R}\}$.}  
\[
\partial_{5}\tilde{F}^{\mu 5}_{\alpha}=\partial_{5}F^{\mu 5}_{\alpha}-2\{\delta_{0}-\delta_{\pi R}\}\omega^{\mu}_{\alpha},
\]     
where $\omega^{\mu}_{\alpha}= \frac{4}{9\sqrt{6}}\kappa e^{-1} \epsilon^{\mu\nu\rho\sigma}C_{\alpha\beta\gamma}F^{\beta}_{\nu\rho}A^{\gamma}_{\sigma}$ are $x^{5}$-independent.  Therefore we have a new field 
\[
\tilde{F}^{\mu 5}_{\alpha}=F^{\mu 5}_{\alpha}-\theta(x^{5})\omega^{\mu}_{\alpha},
\]
which is the upstairs version of~(\ref{bc}).
This field admits a modified vector potential $\tilde{A}^{\mu}_{\alpha}$ with 4D fieldstrengths $\tilde{F}^{\mu\nu}_{\alpha}$ that have a kink at the fixed points, corresponding to the absence of a modified boundary condition.  The modified fields satisfy the Bianchi identity
\[
\partial^{\hat{\mu}}\tilde{F}^{\hat{\nu}\hat{\rho}}+\partial^{\hat{\rho}}\tilde{F}^{\hat{\mu}\hat{\nu}}+\partial^{\hat{\nu}}\tilde{F}^{\hat{\rho}\hat{\mu}}=0.
\]

Now, since the resulting expression for $\partial_{\mu}\tilde{F}^{\mu 5}_{\alpha}$ is invariant under connected abelian transformations, we require that $\tilde{F}^{\mu 5}_{\alpha}$ itself be invariant.  This requires a modification of the local transformation law for the old fields
\[
\tilde{\delta}F^{\mu 5}_{\alpha}=+\theta(x^{5})\delta\omega^{\mu}_{\alpha},
\]
where $\delta\omega^{\mu}_{\alpha}=\frac{4}{9\sqrt{6}}\kappa e^{-1} C_{\alpha\beta\gamma}\epsilon^{\mu\nu\rho\sigma}\partial_{\nu}(F^{\beta}_{\rho\sigma}\alpha^{(\gamma)})$. 

These modifications perhaps look more natural in terms of the 5D Hodge duals of $F^{\hat{\mu}\hat{\nu}}_{\alpha}$: $\mathcal{G}_{\hat{\mu}\hat{\nu}\hat{\rho}}=\frac{e}{2}\epsilon_{\hat{\mu}\hat{\nu}\hat{\rho}\hat{\sigma}\hat{\lambda}}F^{\hat{\sigma}\hat{\lambda}}$.  The dual of $\tilde{F}^{\mu 5}_{\alpha}$ is\footnote{$A_{[\mu_{1}\cdots\mu_{n}]}=\frac{1}{n!}\Sigma A_{\mu_{1}\cdots\mu_{n}}\varepsilon_{\mu_{1}\cdots\mu_{n}}$, where $\varepsilon$ is the permutation symbol and the sum is over permutations of index ordering.} 
\[
\tilde{\mathcal{G}}_{\alpha\,\mu\nu\rho}=\mathcal{G}_{\alpha\,\mu\nu\rho}-\theta(x^{5})\frac{3!\,4}{9\sqrt{6}}\kappa\, C_{\alpha\beta\gamma}F^{\beta}_{[\mu\nu}A^{\gamma}_{\rho]},
\]
while the duals $\tilde{\mathcal{G}}_{\alpha\,\mu\nu 5}$ involve modifications with a kink at the fixed points, corresponding to the absence of a modified boundary condition in the downstairs picture. Note that the $\tilde{\mathcal{G}}_{\alpha\,\mu\nu\rho}$ don't admit a rank-2 tensor potential; these are globally defined objects.

The $\tilde{\mathcal{G}}_{\alpha\,\hat{\mu}\hat{\nu}\hat{\rho}}$ satisfy modified Bianchi identities 
\begin{equation}\begin{split}
d\tilde{\mathcal{G}}_{\alpha\,\mu\nu\rho\sigma}&=-\theta(x^{5})\frac{4!\,2}{9\sqrt{6}}\kappa\, 
C_{\alpha\beta\gamma}F^{\beta}_{[\mu\nu}F^{\gamma}_{\rho\sigma]}\\
d\tilde{\mathcal{G}}_{\alpha\,\mu\nu\rho 5}&=-\{\delta_{0}-\delta_{\pi R}\}\frac{3!\,4}{9\sqrt{6}}\kappa\,
C_{\alpha\beta\gamma}F^{\beta}_{[\mu\nu}A^{\gamma}_{\rho]}.
\end{split}\label{Geqns}\end{equation}
The first relation implies that $\tilde{\mathcal{G}}_{\alpha\,\mu\nu\rho}$ must be invariant under connected abelian transformations, which requires a modified transformation law
\[\tilde{\delta}\mathcal{G}_{\alpha\,\mu\nu\rho}=\theta(x^{5})\frac{3!\,4}{9\sqrt{6}}\kappa\, C_{\alpha\beta\gamma}\partial_{[\mu}(F^{\beta}_{\nu\rho]}\alpha^{(\gamma)}).\] 
Then the second relation is automatically invariant with the original transformation law $\delta \mathcal{G}_{\alpha\,\mu\nu 5}=0$.

Expressions~(\ref{Geqns}) are components of the 4-form Bianchi identities $d\tilde{\mathcal{G}}_{\alpha}$ of the 3-form fieldstrengths $\tilde{\mathcal{G}}_{\alpha}$, and therefore both sides of each expression must be cohomologically trivial.  For closed 4-chains $C_{4}$ in the 5D spacetime not containing the $S^{1}$ covering space of $S^{1}/\mathbb{Z}_{2}$, the first set of conditions require vanishing instanton number 
\[C_{\alpha\beta\gamma}\int_{C_{4}}F^{\beta}\wedge F^{\gamma}=0,\]  
where the wedge product is 4-dimensional.
For closed 4-chains $C_{4}\simeq C_{3}\times S^{1}$, the second set of relations require winding numbers to match
\[C_{\alpha\beta\gamma}\left\{\int_{C_{3}}F^{\beta}\wedge A^{\gamma}-\int_{C'_{3}}F^{\beta}\wedge A^{\gamma}\right\}=0,\]   
where $C_{3}$,$C'_{3}$ are 3-chains on the fixed planes $x^{5}=\{0\},\{\pi R\}$ and the wedge product is again 4-dimensional.  In the downstairs picture, both of these conditions are restrictions on the field configurations on the boundaries.  The 3-forms $\mathcal{G}_{\alpha}$ are globally exact, though on $y=0$ and $y=\pi R$ is not; in other words, an exact 3-form interpolates between the field configurations on the boundaries. 

The case of a YMESGT with \textit{non-abelian} $K_{\alpha}$ involves similar manipulations with the additional coupling term as in~(\ref{nonabelianbc}).  The Bianchi identities are then 
\[\begin{split}
d\tilde{\mathcal{G}}_{\alpha\,\mu\nu\rho\sigma}&=-\theta(x^{5})\frac{4!\,2}{9\sqrt{6}}\frac{\kappa}{g}\, 
C_{\alpha\beta\gamma}\left\{F^{\beta}_{[\mu\nu}F^{\gamma}_{\rho\sigma]}+\frac{9}{4}F^{\beta}_{[\mu\nu}(A^{\alpha'}_{\rho}f^{\gamma}_{\alpha'\beta'}A^{\beta'}_{\sigma]})\right\}\\
d\tilde{\mathcal{G}}_{\alpha\,\mu\nu\rho 5}&=-\{\delta_{0}-\delta_{\pi R}\}\frac{3!\,4}{9\sqrt{6}}\frac{\kappa}{g}\,
C_{\alpha\beta\gamma}\left\{F^{\beta}_{[\mu\nu}A^{\gamma}_{\rho]}+\frac{3}{4}A^{\beta}_{[\mu}(A^{\alpha'}_{\nu}f^{\gamma}_{\alpha'\beta'}A^{\beta'}_{\rho]})\right\},
\end{split}\]
where we've rescaled $gA^{I}_{\hat{\mu}}\rightarrow A^{I}_{\hat{\mu}}$.  The triviality of cohomology takes on the obvious extensions. 
In the first expression, the deviation of the right hand side from being 
$\mathcal{F}\mathcal{F}$ is $\frac{1}{4}F[A,A]$.  In the second expression, 
the deviation of the right hand side from being a Chern-Simons analog is $\frac{1}{12}A[A,A]$.  However, the 4-forms defined from them are in 
the same cohomology class as the 2nd Chern class analogs $C_{\alpha\beta\gamma}\mathcal{F}^{\beta}\mathcal{F}^{\gamma}$ (they are related by an exact 4-form $dQ_{3}$, 
where $Q_{3}$ is known as a `transgression 3-form').     

Five-dimensional sugra has 0- and 1-brane solutions (for simple sugra, see e.g.~\cite{Mizoguchi:1998wv}), and are analogs of the 2- and 5-branes of 11D sugra. Once $p$-branes are added, the above Bianchi identities and cohomology conditions will be modified.  Boundary conditions for the $F^{\mu 5}_{\alpha}$ can arise from magnetic 1-brane sources, with $x^5$ one of the transverse directions to the worldsheets.  The modified Bianchi identity for the $\tilde{\mathcal{G}}_{\alpha}$ indicates that these worldsheets couple to potentials with composite 3-form fieldstrengths of the form $C_{\alpha\beta\gamma}F^{\beta}\wedge A^{\gamma}$ (the product being 4D).  The dual objects in five dimensions are 0-branes whose worldlines are transverse to the boundaries; when they intersect them they result in (-1)-branes (instantons).   If an 11D origin of a particular 5D theory exists via a Calabi-Yau compactification, 0-branes can arise from 2-branes wrapping 2-cycles, and magnetic 1-branes can arise from magnetic 5-branes wrapping 4-cycles.  These cycles, in turn, govern the vector-coupled supergravity sector in 5D since the $h^{I}$ and $h_{I}$ scalar functions are rescalings of Calabi-Yau 2- and 4-cycle volumina, respectively.  

\subsection{The supersymmetry algebra}\label{sec:susyalgebra}
The boundary conditions~(\ref{bcs}) with $N\neq 0$ correspond to odd fermions having $\aleph\neq 0$ in~(\ref{fieldexp}).  In this case, there are then $\delta$-function terms in the Lagrangian and susy transformations.  
The susy transformations close into the usual $\mathcal{N}=2$ superalgebra in the bulk, and should close into an $\mathcal{N}=1$ superalgebra at the fixed points after imposing the fixed-point conditions corresponding to the boundary conditions~(\ref{bcs}).  The relevant transformations to be checked are
\[\begin{split}
\delta^{(2)}_{\epsilon}\delta^{(1)}_{\epsilon}\Psi_{5\,i}&\sim -\frac{i\kappa}{4}\Gamma^{\mu}\partial_{5}\left(\bar{\Psi}^{j}_{\mu}\epsilon^{(2)}_{j}\right)\epsilon^{(1)}_{i}\\
\delta^{(2)}_{\epsilon}\delta^{(1)}_{\epsilon}e^{m}_{5}&\sim -\frac{1}{2}\left(\delta_{(2)}\bar{\Psi}^{i}_{5}\right)\Gamma^{m}\epsilon^{(1)}_{i}\\
\delta^{(2)}_{\epsilon}\delta^{(1)}_{\epsilon}A^{a}_{5}&\sim \frac{i\sqrt{6}}{4}h^{a}\left(\delta_{(2)}\bar{\Psi}^{i}_{5}\right)\epsilon^{(1)}_{i}.
\end{split}\]
The first expression is independent of matter fields, and has the same form as in simple sugra~\cite{JB1}; the terms with a $\delta$-function have no support on $S^{1}/\mathbb{Z}_{2}$ (the closure is then in the weak sense).  In the second and third expressions, there is a fixed-point localized term due to the variation $\delta\Psi^{i}_{5}\sim \partial_{5}\epsilon^{i}$.  But the fixed-point conditions that correspond to~(\ref{bcs}) with $N\neq 0$ imply that two 4-component susy spinors must satisfy the Killing equations everywhere, including at the fixed points where they are degenerate degrees of freedom.  Therefore, as in~\cite{JB1}, we must define a modified transformation $\tilde{\delta}_{\epsilon} \bar{\Psi}^{i}_{5}$ such that  
\begin{equation}
\tilde{\delta}_{\epsilon} \beta_{5}=\delta_{\epsilon}\beta_{5}-2\aleph\{\delta_{0}-\delta_{\pi R}\}\zeta_{\theta}, \label{modifiedsusy}
\end{equation}
along with the conjugate expression for $\beta^{*}_{5}$ (see~(\ref{fieldexp}) for definitions in the 2nd term).  These $\delta$-function issues are, of course, not present when $\aleph=0 \Leftrightarrow N=0$.  

Since the original action in the upstairs picture was invariant with the original susy transformations, we'll have to modify this as well.  At least up to four-fermion terms, the cancellation of $\delta(x^{5})$ contributions in the original variation of the action involves 
\begin{equation}\begin{split}
e^{-1}\delta \mathcal{L}\sim &-\bar{\Psi}^{i}_{\mu}\Gamma^{\mu\nu5}\partial_{\nu}(\delta \Psi_{5\,i})+\frac{1}{2}\bar{\Psi}^{i}_{\mu}\Gamma^{\mu\nu5}\partial_{5}(\delta \Psi_{\nu\,i})+\frac{1}{2}(\delta \bar{\Psi}^{i}_{\mu})\Gamma^{\mu\nu5}\partial_{5}\Psi_{\nu\,i}\\
&\cong  \frac{i}{4\sqrt{6}}h_{I}\bar{\Psi}^{i}_{\mu}\Gamma^{\mu\nu5}(\Gamma^{\hat{\rho}\hat{\sigma}}_{\nu}-4\delta^{\hat{\rho}}_{\nu}\Gamma^{\hat{\sigma}})F^{I}_{\hat{\rho}\hat{\sigma}}\partial_{5}\epsilon_{i},
\end{split}\label{deltasusy}\end{equation}
where $\cong$ is equivalence via integration by parts.  This cancels the contribution of the $\overline{\Psi}_{\mu}F_{\nu\rho}\,\delta_{\epsilon}\Psi_{5}$ term in the variation of the Lagrangian~(\ref{lagrangian}).

Therefore, if we effectively remove the $\delta$-function contribution appearing in $\delta_{\epsilon}\Psi^{i}_{5}$ by defining~(\ref{modifiedsusy}), the uncancelled terms in~(\ref{deltasusy}) must be compensated by adding the following term to the upstairs action~\cite{JB1}: 
\[
e^{-1}\Delta S=\int_{0}\frac{\aleph}{2}\bar{\Psi}^{i}_{c\,\mu}(0)\Gamma^{\mu\nu 5}\Psi_{\theta\,\nu\,i}(0)-\int_{\pi R}\frac{\aleph}{2}\bar{\Psi}^{i}_{c\,\mu}(\pi R)\Gamma^{\mu\nu 5}\Psi_{\theta\,\nu\,i}(\pi R),
\]
where fields and integrations are on the fixed-planes $x^{5}=\{0\},\{\pi R\}$ as indicated, and subscripts on the fields are defined in~(\ref{fieldexp}).  There may be additional fixed-point terms once the form of the spin connection is considered~\cite{JB1}.      

Without much thought, we can get some insights from the gravitino variation $\delta_{\epsilon}\Psi^{i}_{\hat{\mu}}=D_{\hat{\mu}}\epsilon^{i}+\cdots$ by considering continuous vs. jumping fermionic fields.  If we restrict attention to $C^{0}$ fermions, flat space solutions cannot exist since the vanishing of half the susy spinors at the fixed points implies they would have to vanish everywhere.  This leaves open the possibility of warped solutions.  In such backgrounds with warp factor $e^{A\gamma}$, the $\mathbb{Z}_{2}$ parity implies $\partial_{5}\partial_{5}\gamma\propto \{\delta_{0}-\delta_{\pi R}\}$, which in turn indicates the presence of thin  domain walls (associated with magnetic 3-branes on the fixed planes).  On the other hand, the existence of flat space solutions requires Neumann boundary conditions $\partial_{5}\epsilon_{i}|_{\partial M}=0$ in the downstairs picture, so that $\epsilon_{i}(x,y)|_{\partial M}=\epsilon_{i}(x)$.  That is, we impose the boundary conditions with $N\neq 0$ corresponding to fermions that jump across fixed points.  Supersymmetric backgrounds in the case of simple sugra are considered in~\cite{JB1, Bagger:2002rw, JB2}.

\section{Conclusion}\label{sec:conclusion}

The classical actions of 5D $\mathcal{N}=2$ Maxwell- and Yang-Mills-Einstein supergravity theories are not generally invariant under local bosonic and supersymmetries on a manifold with boundary.  If 5D vector fields have propagating modes on the boundaries, the anomalous contributions come from the ``Chern-Simons" term, and are governed by the form of the rank-3 symmetric $C_{IJK}$ that defines the MESGT one starts with (actually, only the subset $C_{\alpha\beta\gamma}$ matters, where $\alpha$ is the index on the surviving 4D vector fields $A^{\alpha}_{\mu}$).  None of these anomalies are present if the theory has $C_{\alpha\beta\gamma}=0$ or when none of the bulk vectors have propagating modes on the boundaries (i.e., they are all $\mathbb{Z}_{2}$-odd). 

Supersymmetry of the action allows non-trivial boundary conditions for some fields that are odd in the ``upstairs picture", which corresponds to jumping behavior at $\mathbb{Z}_{2}$ fixed points.  
Supersymmetry of the action requires such boundary conditions for the components $F^{\mu 5}_{\alpha}$ of the 5D fieldstrengths.  The Hodge dual fieldstrengths satisfy modified Bianchi identities.  For a YMESGT with non-abelian boundary gauge group $K_{\alpha}$, the supersymmetric boundary conditions and Bianchi identities involve composite fields that deviate from `expected' Chern-Simons analogs (analogs in the sense that they are Lie-algebra valued). Gravitini and susy parameters can also satisfy non-trivial boundary conditions, describing a degeneration of fermionic degrees of freedom.  In the ``upstairs picture", this corresponds to allowing odd components to jump at the fixed points; the closure of the susy algebra relies on vanishing support for the resulting Dirac distributions, which follows as in the simple sugra case~\cite{JB1}. 

In a YMESGT, anomaly inflow of gauge symmetries must be compensated either by coupling bulk $\mathcal{N}=2$ hypermultiplets or boundary $\mathcal{N}=1$ chiral multiplets in $\mathbb{C}$-representations of the surviving boundary gauge group $K_{\alpha}$.  In a MESGT, the anomalies in the local abelian symmetries require the addition of boundary fermions charged with respect to the $A^{\alpha}_{\mu}$ Maxwell fields.  In this sense the MESGT is ``gauged" on the boundaries.  \vspace{2mm}\\  
{\bf Acknowledgement:} \\
Work supported by the European Commission RTN program ``Constituents, Fundamental Forces and Symmetries of the Universe" MRTN-CT-2004-005104 and by INFN, PRIN prot.2005024045-002.
 
\begin{appendix}
\section{Some minor details regarding 5D MESGTs}
We use the 5D spacetime signature $(-++++)$.  The matrices $\Gamma^{\hat{m}}$ with flat spacetime indices are taken as 
\[
\Gamma^{m}=\left(\begin{array}{cc}
0 & \sigma^{m}\\
-\sigma^{m} & 0
\end{array}\right)\;\;\;\;\Gamma^{\bar{5}}=\left(\begin{array}{cc}
i & 0\\
0 & -i
\end{array}\right),
\]
where $\sigma^{m}$ are the Pauli matrices with $\sigma^{\bar{0}}=\mathbb{I}$.  The charge conjugation matrix $C$ is taken to be 
\[
C=\left(\begin{array}{cc}
e & 0\\
0 & -e
\end{array}\right)\;\;\;\;\mbox{where}\;\;\;\;
e=\left(\begin{array}{cc}
0 & -1\\
1 & 0
\end{array}\right),
\]
which satisfies 
\[
C^{T}=-C=C^{-1}\;\;\;\;C\Gamma^{m}C^{-1}=(\Gamma^{m})^{T}.
\]
The gravitini $\Psi^{i}_{\hat{\mu}}$ and spin-1/2 fermions $\lambda^{i}_{\tilde{p}}$ carry $SU(2)_{R}$ index $i=1,2$.  These are known as symplectic-Majorana spinors, which satisfy the relation 
\[
\overline{\lambda}^{i}=\lambda^{i\,T}C.
\]  
The Dirac conjugate is $\overline{\lambda}^{i}=(\lambda_{i})^{\dagger}\Gamma^{0}$ with the raising and lowering convention
\[
\lambda^{i}=\epsilon^{ij}\lambda_{j}\;\;\;\;\lambda_{j}=\lambda^{i}\epsilon_{ij}.
\]
The fermions of the theory can then be written as in~(\ref{spinors1}).

The $h^{I}$ are $n_{V}+1$ functions of the $n_{V}$ scalars $\phi^{\tilde{x}}$ determined by solutions of $\mathcal{V}=C_{IJK}h^{I}h^{J}h^{K}=1$.  The symmetric tensor in the kinetic terms for the vectors and scalars is $\stackrel{\circ}{a}_{IJ}=-\partial_{I}\partial_{J}\mathcal{V}_{1}$.  This raises and lowers indices as in $h_{I}=\stackrel{\circ}{a}_{IJ}h^{J}$.  These functions along with $h^{I}_{\tilde{x}}$ appearing in the theory satisfy the following relations 
\[\begin{split}
&h_{I}=C_{IJK}h^{J}h^{K}\;\;\;\;\;\;\;\;h^{I}h_{I}=1\\
&h^{I}_{\tilde{x}}=-\sqrt{\frac{3}{2}}\frac{\partial h^{I}}{\partial\phi^{\tilde{x}}}\;\;\;\;\;\;\;\;h^{I}_{\tilde{x}}h_{I}=0
\end{split}
\]    
Furthermore, $h^{I}_{\tilde{p}}=f^{\tilde{x}}_{\tilde{p}}h^{I}_{\tilde{x}}$, where $f^{\tilde{x}}_{\tilde{p}}$ are vielbein of the scalar manifold.  
\end{appendix}

\end{document}